# 7.86 kV GaN-on-GaN PN Power Diode with BaTiO₃ for Electrical Field Management

Yibo Xu, Vijay Gopal Thirupakuzi Vangipuram, Vishank Telasara, Junao Cheng, Yuxuan Zhang, Tadao Hashimoto, Edward Letts, Daryl Key, Hongping Zhao, Senior Member, IEEE, and Wu Lu, Senior Member, IEEE

*Abstract* — **Device based on GaN have great potential for high power switching applications due to its high breakdown field and high electron mobility. In this work, we present the device design of a vertical GaN-on-GaN PN power diode using high dielectric constant (high-k) dielectrics for electrical field management and high breakdown voltages, in together with guard-rings and a field plate. The fabricated diodes with a 57 μm thick drift layer demonstrated a breakdown voltage of 7.86 kV on a bulk GaN substrate. The device has an on-resistance of 2.8 mΩ·cm² and a Baliga figure of merit of 22 GW/cm².**

*Index Terms* — **Vertical GaN device, Power Diode, Device Fabrication, MOCVD growth, high-k, BTO**

## I. INTRODUCTION

Gallium nitride (GaN) based devices hold immense potential for high-power applications due to its unique material properties. Development and improvements in GaN growth with low impurity incorporation, and the increased availability of low-defect density native GaN substrates for thick homoepitaxial films, have provided a pathway towards achieving vertical GaN power devices closer to gallium nitride material limits [1][2]. Specifically, to achieve a high breakdown voltage (> 5 kV), it is critical to have a thick drift layer with a low doping density and effective electrical field management [3]. Therefore, a fast growth rate is required in order to increase drift layer thicknesses to further enhance breakdown voltages while also reducing the required time for growth. To ensure devices do not breakdown prematurely, previous work has shown multiple methods to achieve high breakdown voltages with GaN-on-GaN PN diodes in terms of field management schemes, such as guard rings (GRs), junction termination extension (JTE), and field plate. Ohta et al. achieved nearly 5kV breakdown voltage utilizing a device structure that included field plates, spin-on-glass (SOG) as well as guard rings [4] [5]. Yates et al. have demonstrated a ~6kV PN diode utilizing a step etched JTE structure with a 50μm drift layer thickness [6].

Furthermore, the implementation of high-k dielectrics is also used in a variety of electric devices for electrical field managements and have demonstrated excellent performance in radio frequency (RF) devices and power devices [7][8]. Cheng et al. utilized a high-k passivation layer of BiZnNbO (BZN) on GaN HEMTs which increased 44% of the device breakdown [7]. Shankar et al. demonstrated a combination of high-k layer and metal field plate structure that can effectively enhance the device breakdown voltage and reduce the electrical field at the edge of the diode electrode up to 88% [8]. As one of the high-k dielectrics, BTO is also promising in power devices due to its high dielectric constant and high breakdown field. In this work, the implementation of BTO is studied and utilized as one of the electric field mitigation methods. As a result, we successfully demonstrated a PN diode with a breakdown voltage of 7.86 kV on a bulk GaN substrate.

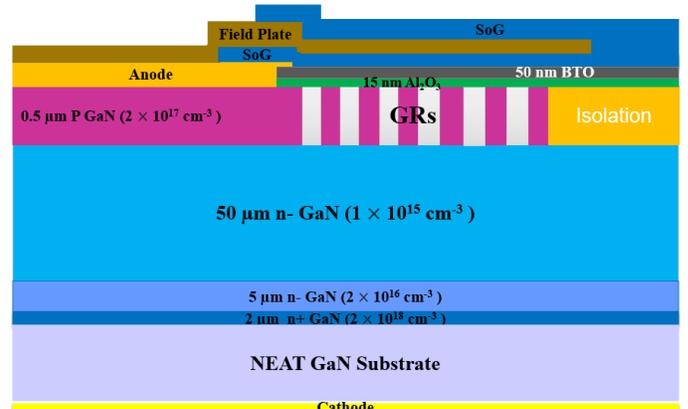

**Fig. 1** Cross-section schematic of the designed device

## II. DEVICE DESIGN

The goal of this work is to demonstrate GaN PN diodes with a breakdown voltage of 7.5 kV. To achieve this goal, a device design with a breakdown voltage of 9 kV is required, considering the breakdown efficiency of ~ 80% [9]. The final designed GaN-on-GaN PN diode structure is shown in Fig. 1.

Numerical simulations were conducted using Silvaco Atlas to analyze the layer structure and study electric field management. The simulated structure has a drift layer thickness of 50 μm with a doping concentration of $1\times10^{15}$ cm$^{-3}$ on top a 5 μm n- GaN with $N_d$-$N_a$ concentration of $2\times10^{16}$ cm$^{-3}$ and a 2 μm n+ GaN of $2\times10^{16}$ cm$^{-3}$. In the simulation, the breakdown field for GaN as 3.5 MV/cm² is considered. GR structures were implemented firstly, the variations on the number of GRs, GR dimensions, and the spacings in between are optimized and simulated. The final design includes 6 GRs with a width of 8 μm and a spacing of 14 μm between each GR. The simulated device breakdown voltage is 8.33 kV with GRs included, while the same device structure without GRs has a breakdown voltage of 7.1 kV. The peak electric field is pushed from the edge of the

Yibo Xu, Vijay Gopal Thirupakuzi Vangipuram, Vishank Telasara, Junao Cheng, Hongping Zhao, Wu Lu are with Department of Electrical and Computer Engineering, The Ohio State University, Columbus, Ohio, 43210, USA (Email: lu.173@osu.edu, zhao.2592@osu.edu)

Tadao Hashimoto, Edward Letts, Daryl Key are with SixPoint Materials, SixPoint Materials, Buellton, California, 93427, USA.

anode to the edge of the last GR. Next, the first layer of SoG is added around and step-on to the anode area, followed by a thick field plate that covers the anode area and extended to the isolation area for 12 µm. The last layer within the simulation is a 3 µm SoG on the top. Figure 2 illustrates the electric field distribution under a reverse bias of 9.05 kV for the device with the above mentioned layer structure. This design makes the electric field at the edges of each GR more uniform and the field peaks at the edge of the field plate.

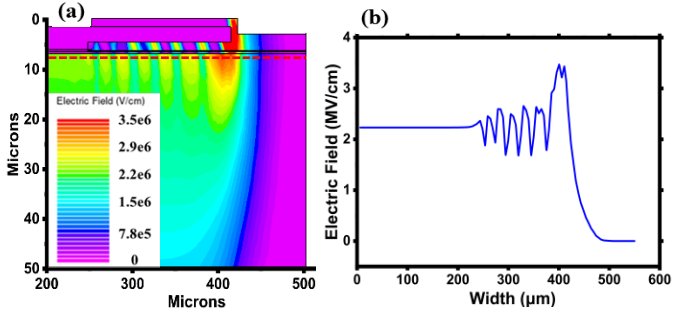

Fig. 2 (a) The enlarged simulated device electric field profile under 9.05 kV reverse bias with GRs, SoG, and field plate. (b) Electric field profile along the cutline starting from the center of the anode (0 micron).

To further mitigate the electric field at the edge of the field plate, a stack of high-k dielectric layers containing a thin layer of $Al_2O_3$ and BTO is added next to the anode area. As a result, the device breakdown voltage is improved to 9.65 kV as shown in Figure 3. The dielectric layers lead to a 600 V increase of breakdown voltage based on the simulation.

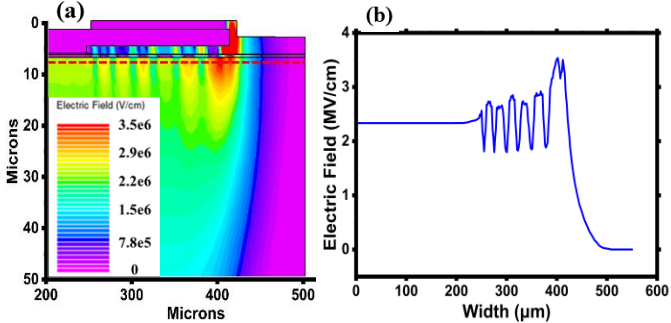

Fig. 3 (a) The enlarged simulated electric field profile under a reverse bias of 9.65 kV with GRs, high-k dielectrics, field plate, and SoG. (b) Electric field profile along the cutline from the center of the anode (0 micron).

### III. MOCVD GROWTH AND FABRICATION

The epitaxial structure was grown on a 2" near equilibrium ammonothermal bulk GaN (NEAT GaN) substrate with a threading dislocation density of ~ $2 \times 10^5$ cm$^{-2}$ [10] in a GaN MOCVD vertical rotating-disk reactor. For the detailed growth process is described in [9, 11]. Briefly, trimethylgallium (TMGa) and ammonia were used as precursors for GaN growth while silane ($SiH_4$) and bis(cyclopentadienyl)magnesium ($Cp_2Mg$) were used as sources for Si and Mg dopants respectively for n-type and p-type doping. The MOCVD growth starts with a ~2 µm n$^+$ GaN layer with a targeted Si concentration of $2 \times 10^{18}$ cm$^{-3}$ followed by a 5 µm n$^-$ GaN layer with a $N_d$-$N_a$ concentration of $2 \times 10^{16}$ cm$^{-3}$ at a growth rate of ~5 µm/hour. Then, the 57 µm thick drift layer was grown at a high growth rate (~11.4 µm/hour) with an average $N_d$-$N_a$ concentration of $1.5 \times 10^{15}$ cm$^{-3}$ in regions of interest. This was followed by a 500 nm p+ GaN layer with an expected hole concentration of $2 \times 10^{17}$ cm$^{-3}$ and a 20 nm p++ GaN cap layer for ohmic contacts. The growth was done at 1010°C at 200 torr pressure utilizing $H_2$ as a carrier gas and was followed by an in-situ thermal anneal for p-type activation for 30 min at 824°C in an $N_2$ ambient. The $N_d$-$N_a$ concentrations for the individual layers presented were initially validated through C-V analysis on separate calibration samples grown prior to the full structure and were further confirmed for the drift layer through C-V analysis done on the actual fabricated devices. The final drift layer thickness of 57 µm was confirmed by cross-sectional cathodoluminescence imaging.

For device fabrication, Ni/Pt/Au and Ti/Al/Ni/Au metal stacks were used for p-contact and n-contact, respectively. After rapid thermal annealing, the optimized p-contact resistance based on TLM measurement is $0.189$ $m\Omega \cdot cm^2$. GRs were implemented by nitrogen ion implantation at five different energies and doses respectively of 380 keV, 230 keV, 115 keV, 45 keV, 12.5 keV and doses of $3.2 \times 10^{13}$, $9.8 \times 10^{12}$, $6.7 \times 10^{12}$, $5 \times 10^{12}$, $1 \times 10^{13}$ in unit of ion/cm$^2$ [9]. 15 nm of $Al_2O_3$ was deposited by atomic layer deposition (ALD), followed by 50 nm of BTO by sputter deposition at 670 °C. An ICP/RIE dry etch with $BCl_3$ was used to etch off the BTO around anode area. After p-contact metal deposition, the first layer of 700 nm thick SoG was applied on top of the sample. Dry etching with $SF_6$ was used to etch off the SoG to open up the anode area. A 50nm/900nm thick of Ti/Al metal stack was deposited as the field plate which covered the complete anode area and further extended up until the isolation area for electric field mitigation. Finally, a 3 µm thick SoG was applied on top for surface passivation. The device has a circular active region of 200 µm in diameter.

### IV. DEVICE PERFORMANCE AND DISCUSSION

The C-V measurements were conducted on the fabricated device with a reverse bias range from 0V to -10 V under room temperature and a built-in voltage of 4.16 V was extracted as shown in Fig. 4 (a). The actual drift layer doping profile along with depletion width was also extracted accordingly. Fig. 4(b) shows a $1 \sim 2 \times 10^{15}$ cm$^{-3}$ electron concentration within the device drift layer.

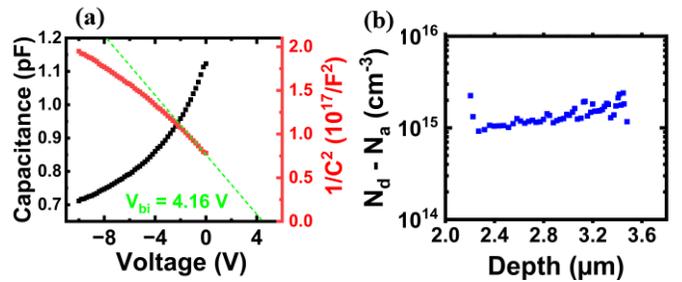

Fig. 4 (a) Device C-V and $\frac{1}{C^2}$ profile under a reverse bias from 0 V to -10 V. (b) Drift layer doping concentration extracted from C-V measurements.

The forward and reverse characteristics have been performed as shown in Fig. 5. All measurements were performed at room temperature. The forward characteristic is shown in Fig. 5(a), the device displayed a turn-on voltage of 4.55 V at 100 A/cm$^2$ and an on-resistance of 2.8 mΩ·cm$^2$. By subtracting the contact resistance from the device on-resistance, the drift layer resistance was estimated to be 2.61 mΩ·cm$^2$ and the mobility of 1243 cm$^2$/Vs was calculated assuming a doping concentration of $1.5 \times 10^{15}$ cm$^{-3}$. Since the device obtains a high breakdown, thus, a high-voltage power supply was employed for breakdown measurement, and the anode surface was covered by Fluorinert during breakdown measurement to prevent air breakdown. The reverse and breakdown characteristics are presented in Fig. 5(b). The device exhibited a low leakage current density of $1.7 \times 10^{-7}$ A/cm$^2$ under -200 V. The breakdown voltage of the device reached 7.86 kV at a current density of 2 mA/cm$^2$.

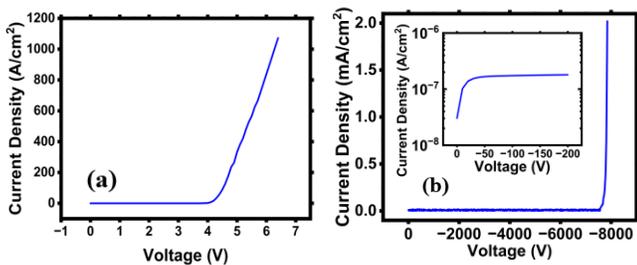

**Fig. 5** (a) Device forward characteristics. (b) Device breakdown behavior at 7.86 kV, and leakage current at reverse bias of 200 V.

Based on the breakdown voltage and on-resistance, the BFOM was calculated to be 22 GW/cm$^2$. Figure 6 shows how this device is benchmarked with reported devices with breakdown voltages larger than 3 kV and BFOM larger than 5 GW/cm$^2$. The obtained device breakdown voltage of 7.86 kV is the highest breakdown voltage reported to date in literature for GaN vertical pn-didoes. The breakdown voltage efficiency obtained for the fabricated device is 81% compared to numerical simulation. The breakdown voltage is increased by 35% and 6.6% by comparing diode with and without field management and with and without high-k implementation respectively. We attribute the high breakdown voltage and BFOM demonstrated here to the NEAT GaN substrate with a low threading dislocation density, the high material quality from epitaxial growth, along with the optimized device design and fabrication processes. Devices with higher breakdown voltage ratings can be expected using this approach with a thicker drift layer.

## V. CONCLUSION

In this work, the design, fabrication, and characterization of a GaN-on-GaN vertical power p-n diode with a 57 μm thick drift layer. The device demonstrated a breakthrough breakdown voltage of 7.86 kV. This represents the highest breakdown voltage reported for a GaN-based power p-n diode to date. Additionally, the device exhibited a low on-resistance and a BFOM of 22 GW/cm$^2$. The outstanding device performance is the result of the combination of high-quality material, high-k dielectric implementation, along with a series of optimized electric field management techniques.

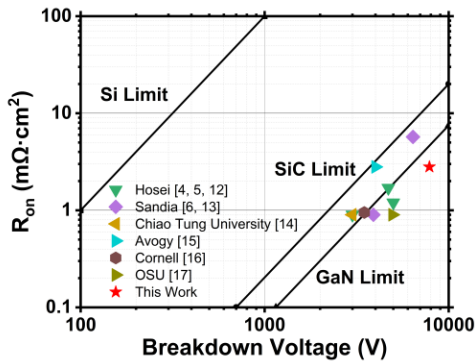

**Fig. 6** The location of this work presenting in Baliga Figure of Merit compare with other reported works.

## Reference

[1] T. Oka, "Recent development of vertical GaN power devices," Japanese Journal of Applied Physics, vol. 58, no. SB, pp. SB0805, Apr. 2019, doi: 10.7567/1347-4065/ab02e7.

[2]: K. Hiramatsu, H. Miyake, K. Fujito, Y. Mitani, and T. Fukushima, "High-purity GaN epitaxial layers for power devices on low-dislocation-density GaN substrates," IEEE Transactions on Electron Devices, vol. 54, no. 11, pp. 3043-3048, Nov. 2007. DOI: 10.1109/TED.2007.907322.

[3] R. J. Kaplar, A. T. Binder, L. Yates, A. A. Allerman, A. M. Armstrong, M. H. Crawford, J. R. Dickerson, C. E. Glaser, J. Steinfeldt, V. M. Abate, M. L. Smith, G. Pickrell, P. Sharps, J. D. Flicker, J. Neely, L. Rashkin, L. Gill, T. Monson, J. Bock, G. Subramania, E., and J. A. Cooper, "Development of vertical gallium nitride power devices for use in electric vehicle drivetrains", 240th ECS Meeting (Digital Meeting), Oct. 14, 2021, doi: 10.1149/ma2021-02341001mtgabs.

[4]: 1 H. Ohta, N. Kaneda, F. Horikiri, Y. Narita, T. Yoshida, T. Mishima, and T. Nakamura, "Vertical GaN pn junction diodes with high breakdown voltages over 4 kV." IEEE Electr. Device L., vol. 36, pp. 1180-1182, Nov. 2015, doi: 10.1109/LED.2015.2478907.

[5]: H. Ohta, K. Hayashi, F. Horikiri, M. Yoshino, T. Nakamura, and T. Mishima. "5.0 kV breakdown-voltage vertical GaN p–n junction diodes." Jpn. J. Appl. Phys., vol. 57, p. 04FG09, Feb 2018, doi:10.7567/ssdm.2017.n-6-02.

[6]: A. L. Yates, B.P. Gunning, M. H. Crawford, J. Steinfeldt, M. L. Smith, V. M. Abate, J. R. Dickerson, A. Binder, A. A. Allerman, A. M. Armstrong,and R. J. Kaplar, "Demonstration of 6.0-kV Breakdown Voltage in Large Area Vertical GaN p-n Diodes With Step-Etched Junction Termination Extensions", IEEE Trans. On Electron Devices, Vol 69, No. 4, April, 2022, doi: 10.1109/TED.2022.3154665.

[7]: J. Cheng, M. W. Rahman, A. Xie, H. Xue, S. H. Sohel, E. Beam, C. Lee, H. Yang, C. Wang, Y. Cao, S. Rajan, and W. Lu, "Breakdown Voltage Enhancement in ScAlN/GaN High-Electron-Mobility Transistors by High-k Bismuth Zinc Niobate Oxide," in IEEE Transactions on Electron Devices, vol. 68, no. 7, pp. 3333-3338, July 2021, doi: 10.1109/TED.2021.3084136.

[8]: B. Shankar, S. K. Gupta, W. R. Taube and J. Akhtar, "High-k dielectrics based field plate edge termination engineering in 4H-SiC Schottky diode," in Semiconductor Science and Technology, vol. 32, no. 4, p. 045016, 2017, doi: 10.1088/1361-6641/aa5b5e.

[9] V. Talesara, Y. Zhang, Z. Chen, H. Zhao and W. Lu, "Design and development of 1.5 kV vertical GaN pn diodes on HVPE substrate," in IEEE Journal of Emerging and Selected Topics in Power Electronics, vol. 7, no. 3, pp. 1597-1605, Sept. 2019, doi: 10.1109/JESTPE.2019.2915212.

[10] T. Hashimoto, E. R. Letts and D. Key, "Progress in Near-Equilibrium Ammonothermal (NEAT) growth of GaN substrates for GaN-on-GaN semiconductor devices," in Crystals, vol. 12, no. 8, p. 1085, Aug. 2022, doi: 10.3390/cryst12081085.

[11] Y. Zhang, Z. Chen, W. Li, A. R. Arehart, S. A. Ringel, and H. Zhao, "Metalorganic chemical vapor deposition gallium nitride with fast growth rate for vertical power device applications," Phys. Status Solidi A, vol. 218, no. 6, Jan. 2021, doi: 10.1002/pssa.202000469.

[12]: Y. Hatakeyama, K. Nomoto, A. Terano, N. Kaneda, T. Tsuchiya, T. Mishima and T. Nakamura, "High-Breakdown-Voltage and LowSpecific-on-Resistance GaN p–n Junction Diodes on Free-Standing GaN Substrates

Fabricated Through Low-Damage Field-Plate Process", Jpn. J. Appl. Phys. vol. 52 pp. 028007, Feb 2013 doi: 10.7567/jjap.52.028007.

[13]: A. M. Armstrong, A. A. Allerman, A. J. Fischer, M. P. King, M. S. Van Heukelom, M. W. Moseley, R. J. Kaplar, J. J. Wierer, M. H. Crawford, and J. R. Dickerson. "High voltage and high current density vertical GaN power diodes." Electron. Lett., vol. 52, p. 1170-1171, April 2016, doi: 10.1049/el.2016.1156.

[14]: S.-W. H. Chen, H.-Y. Wang, C. Hu, Y. Chen, H. Wang, J. Wang, W. He, X. Sun, H.-C. Chiu, H.-C. Kuo, W. Wang, K. Xu, D. Li, X. Liu, "Vertical GaN-on-GaN PIN diodes fabricated on free-standing GaN wafer using an ammonothermal method", Journal of Alloys and Compounds, vol. 804, pp. 435-440, July. 2019, doi: 10.1016/j.jallcom.2019.07.021.

[15] I. C. Kizilyalli, T. Prunty, and O. Aktas. "4-kV and 2.8 mΩ-cm$^2$ vertical GaN pn diodes with low leakage currents." IEEE Electron Device L., vol. 36, p.1073-1075, Oct. 2015, doi: 10.1109/LED.2015.2474817.

[16] K. Nomoto, Z. Hu, B. Song, M. Zhu, M. Qi, R. Yan, V. Protasenko, "GaN-on-GaN pn power diodes with 3.48 kV and 0.95 mΩ-cm2: A record high figure-of-merit of 12.8 GW/cm2." 2015 IEEE International Electron Devices Meeting (IEDM), 2015, pp. 9.7.1-9.7.4, doi: 10.1109/IEDM.2015.7409665.

[17] V. Telasara, Y. Zhang, V. Vangipuram, H. Zhao, W. Lu, "Vertical GaN-on-GaN pn power diodes with Baliga figure of merit of 27 GW/cm$^2$." Appl. Phys. Lett. Vol.122, no. 12, Mar. 2023, doi: 10.1063/5.0135313.